\documentclass[12pt]{article} 
\usepackage{amsmath,natbib,rotating,xcolor,url}
\usepackage[a4paper,total={6.0in, 9.0in}]{geometry}

\renewcommand{\baselinestretch}{\originalbaselinestretch}

\title{%
\textbf{A goodness-of-fit test for the Birnbaum-Saunders distribution based on the probability plot}}
\author{
\textbf{Chanseok Park} \\
Applied Statistics Laboratory\\
Department of Industrial Engineering\\ Pusan National University\\
Busan, Republic of Korea
\and
\textbf{Min Wang}\thanks{Corresponding author. Email: min.wang3@utsa.edu} \\
Department of Management Science and Statistics \\
The University of Texas at San Antonio \\
San Antonio, TX, USA
}
\date{}

\begin{document}
\maketitle
\begin{abstract}
In the present paper, we develop a new goodness-of-fit test for the Birnbaum-Saunders distribution based on the probability plot. 
We utilize the sample correlation coefficient from the Birnbaum-Saunders probability plot as a measure of goodness of fit.
Unfortunately, it is impossible or extremely difficult to obtain an explicit distribution of this sample correlation coefficient. To address this challenge, we employ extensive Monte Carlo simulations to obtain the empirical distribution of the sample correlation coefficient from the Birnbaum-Saunders probability plot. This empirical distribution allows us to determine the critical values alongside their corresponding significance levels, thus facilitating the computation of the $p$-value when the sample correlation coefficient is obtained. Finally, two real-data examples are provided for illustrative purposes. 
\end{abstract}

\section{Introduction}
When analyzing a set of experimental observations and building a statistical model, it is necessary to perform goodness-of-fit tests for any underlying probability distribution. Probability plotting methods \citep{Wilk/Gnanadesikan:1968} are commonly adopted for constructing goodness-of-fit tests, as these probability plots serve as intuitive visualization tools in data analysis. Visual assessments can be quite subjective in practical applications, such as the Q-Q plot for normality and the Weibull plot \citep{Nelson:1982} for the Weibullness of data. 


It is well-known that one can utilize the sample correlation coefficient from
the probability plot to construct formal goodness-of-fit tests for normality and
Weibullness, respectively. For the normality test, the methods using the sample correlation
coefficient from the Q-Q plot are advocated by \cite{Mage:1982},
\cite{Looney/Gulledge:1985}, \cite{Vogel:1986}, \cite{Neter/etc:1996},
\citet[][\S 3.5]{Kutner/etc:2005},
\citet[][\S 2.4]{Montgomery:2005}, \citet[][\S 4.6]{Johnson/Wichern:2007},
among others. For the Weibullness test,  the goodness-of-fit tests based on the sample correlation coefficient from the Weibull plot were recently developed by \cite{Park:2017b,Park:2018c}. 

Building upon the concepts of normality and Weibullness discussed earlier, a logical progression involves the creation of a goodness-of-fit test tailored to the Birnbaum-Saunders distribution proposed by \cite{Birnbaum/Saunders:1969a} for modeling the fatigue failure caused under cyclic loading. This type of the goodness-of-fit test harnesses the sample correlation coefficient extracted from the Birnbaum-Saunders probability plot. The challenge in this development lies in determining the explicit distribution of the sample correlation coefficient from the Birnbaum-Saunders probability plot. This is crucial for constructing a critical region for the proposed goodness-of-fit test. Unfortunately, attaining the explicit distribution of the sample correlation coefficient might prove to be either impossible or exceedingly complex. Analogous to
\cite{Filliben:1975}, \cite{Looney/Gulledge:1985} and \cite{Park:2017b}, we may employ the Monte Carlo simulation method to approximate the empirical distribution of the sample correlation coefficient
from the Birnbaum-Saunders probability plot. As emphasized by \cite{Evans/Johnson/Green:1989},  studies like the Monte-Carlo study carried out by \cite{Shapiro/Wilk/Chen:1968} have consistently shown that the Shapiro-Wilk statistic exhibits superior power in detecting deviations from normal distributions compared to other statistics. This suggests its ability to identify data originating from a wide range of alternative distributions.

An argument could be made that the Kolmogorov-Smirnov test can serve as a goodness-of-fit test for the Birnbaum-Saunders distribution. However, it is important to note that the most significant limitation of this test is its requirement for the distribution to be completely specified. In other words,  this implies that in order to apply the test, the true values of the population parameters must be known, which are however seldom occurred in most practical cases. One may refer to \cite{NIST-1.3.5.16-2018},
Section 10.2 of \cite{Lawless:2003}, and the \texttt{ks.test()} function in the R language \citep{R:2023} for more details. Unlike the Kolmogorov-Smirnov test, the proposed method is invariant with respect to the population parameters. As a result, it does not require knowledge of the population parameter values to conduct the goodness-of-fit test for the Birnbaum-Saunders distribution.

The rest of the paper is organized as follows.  In Section~\ref{section:02}, we briefly
introduce the Birnbaum-Saunders probability plot and a method for
constructing the formal goodness-of-fit test for the Birnbaum-Saunders distribution. 
The real-data examples are provided in Section~\ref{section:03}.  
This papaer ends with concluding
remarks in Section~\ref{section:04}, with figures of empirical cumulative distribution function and tables of the critical values deferred to the Appendix. 

\section{{The Birnbaum-Saunders distribution and linearized probability plot}} \label{section:02}
In this section, we provide  a brief introduction to the Birnbaum-Saunders distribution proposed by \cite{Birnbaum/Saunders:1969a} and the probability plot proposed by \cite{Chang/Tan:1994b}. 

The Birnbaum-Saunders distribution originally derived a fatigue life distribution based on 
a physical fatigue process with respect to crack growth which causes failure.
The probability density function (pdf) and 
the cumulative distribution function (CDF) 
of the Birnbaum-Saunders distribution are given by
\begin{align}
f(t) &= {\frac{1}{2{\alpha}{\beta}\sqrt{2{\pi}}}}
 \Bigg[\sqrt{\frac{\beta}{t}}+\Bigg({\frac{\beta}{t}}\Bigg)^{3/2}\Bigg]
  \exp\Bigg[{ -\frac{1}{2{\alpha}^2}}
 \Bigg(\frac{t}{\beta}-2+{\frac{\beta}{t}}\Bigg)\Bigg] \notag \\
\intertext{and}
F(t) &= \Phi \Bigg[ \frac{1}{\alpha}  \label{EQ:CDF}
                    \Bigg( \sqrt{\frac{t}{\beta}}  - \sqrt{\frac{\beta}{t}} \Bigg) \Bigg], \quad t>0,
\end{align}
respectively, where $\alpha>0$ is the shaper parameter, $\beta>0$ is the scale parameter, and $\Phi(\cdot)$ is the standard normal CDF. 

Denote the ordered observations $t_i$ by $t_{(i)}$. 
Using an empirical estimate of $F(t_{(i)})$ in (\ref{EQ:CDF}) denoted by $p_i = \hat{F}(t_{(i)})$,
we set 
\[
p_i = \Phi \Bigg[ \frac{1}{\alpha}  
                  \Bigg( \sqrt{\frac{t_{(i)}}{\beta}}  - \sqrt{\frac{\beta}{t_{(i)}}} \Bigg) \Bigg] . 
\]
Several versions of the estimate $p_i = \hat{F}(t_{(i)})$ have been suggested in the statistics literature,  but the most popular one is the method by \cite{Blom:1958} and \cite{Wilk/Gnanadesikan:1968}, which is given by 
$p_i = (i-1/2)/n$ for $n\ge 11$ and $p_i = (i-3/8)/(n+1/4)$ for $n \le 10$.

It can be easily shown that the CDF of the Birnbaum-Saunders distribution can be linearized as
\[
v_i = -\frac{\sqrt{\beta}}{\alpha} + \frac{1}{\alpha\sqrt{\beta}}\cdot u_i,
\]
where  $u_i = t_{(i)}$ and $v_i = \sqrt{t_{(i)}} \Phi^{-1}(p_i)$. By letting $\bar{u}= \frac{1}{n} \sum_{i=1}^{n}u_i$ and  $\bar{v}= \frac{1}{n} \sum v_i$, we can obtain the sample correlation coefficient of $(u_i, v_i)$ given by 
\begin{equation}
R = \frac{\displaystyle\sum_{i=1}^{n}(u_i -\bar{u})(v_i -\bar{v})}%
{\sqrt{\displaystyle\sum_{i=1}^{n}(u_i -\bar{u})^2 \cdot\sum_{i=1}^{n}(v_i-\bar{v})^2}},
\label{EQ:correlation}
\end{equation}
which is invariant with respect to the population parameters ($\alpha$ and $\beta$);  see, for example,  \cite{Vogel:1986} and \cite{Vogel/Kroll:1989}. This invariance property eliminates the need to estimate the population parameters in order to construct a formal statistical hypothesis test using the sample correlation coefficient for the hypotheses of the form 
\begin{center}
$H_0$: Birnbaum-Saunders distribution  \emph{~versus~}  $H_1$: non-Birnbaum-Saunders.
\end{center}
Using the sample correlation coefficient from the Birnbaum-Saunders probability plot as a test statistic, we could construct a critical region which rejects the null hypothesis. 
That is, we reject $H_0$ when $R < r_\gamma$, where $r_\gamma$ is the associate critical value corresponding at a given significance level $\gamma$. However, a significant challenge arises in determining a critical region for the Birnbaum-Saunders goodness-of-fit test under the null hypothesis. As far as our current knowledge goes, the distribution of the sample correlation coefficient from the Birnbaum-Saunders probability plot under the null hypothesis remains unknown. Consequently, the empirical distribution needs to be established through the use of Monte Carlo simulations.

\begin{figure}[t!]
\centering\includegraphics[width=6in]{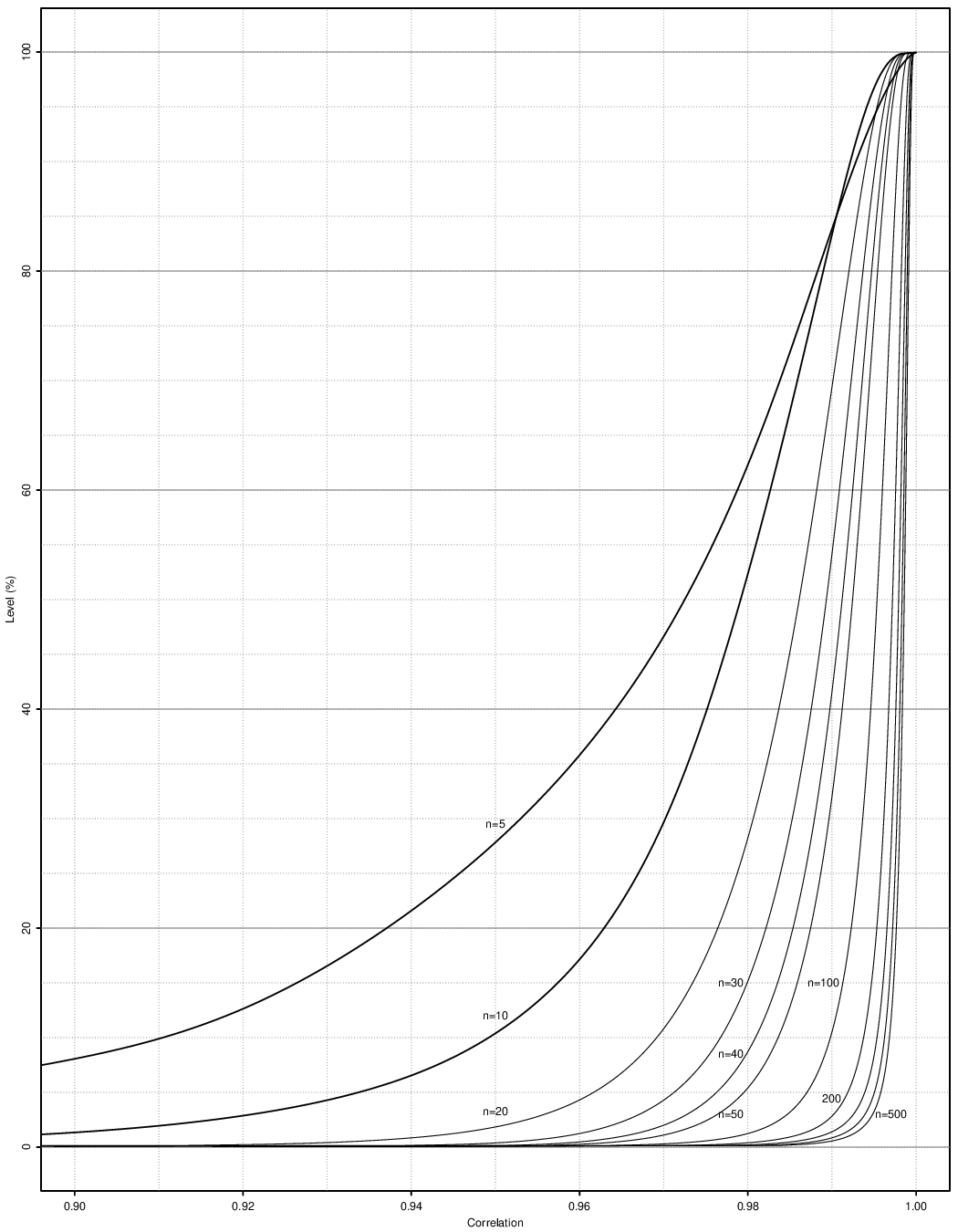}  
\caption{The empirical CDF.\label{FIG:plot1}}
\end{figure}

To obtain the empirical distribution,  we computed one hundred million $I=10^8$ sample correlation coefficients from the Birnbaum-Saunders probability plot with a random sample of size $n$ from the Birnbaum-Saunders distribution. We repeated this process each of $n=3,4,\ldots,1000$ and obtained the empirical CDF of the sample correlation coefficient from the Birnbaum-Saunders probability plot. Then by inverting this empirical CDF at the significance level of $\gamma$,
we obtained the empirical critical value at a given significance level. We provided these critical values in Tables~\ref{TBL:criticalvalues1} and \ref{TBL:criticalvalues2} at the significance levels of
0.5\%, 1\%, 2\%, 2.5\%, 5\%, 7.5\%, 10\%, 12.5\%, 15\%, 20\%, 25\%, and  50\% 
for each of sample sizes, $n=3,4,\ldots,100$ and $n=110,120,\ldots,500, 550, \ldots, 1000$.  In addition, a subset of these results is plotted in Figure~\ref{FIG:plot1}. In practice, as one is generally interested in the critical values at which the significance levels are
less than 10\%, we zoomed in Figure~\ref{FIG:plot1} including only the levels up to 10\%, shown in Figure~\ref{FIG:plot2}. We believe that these values along with the figures are sufficient for most practical problems.

It is worthwhile to discuss the accuracy of the critical values obtained using
Monte Carlo simulations.
Let $\hat{F}^{-1}_I(p)$ be the empirical quantile of $p$ obtained from the $I$ iterations 
and $F^{-1}(p)$ be the true quantile of $p$. Then it is easily seen from
Corollary~21.5 of \cite{Vaart:1998} that the sequence
$\sqrt{I}\big(\hat{F}_I^{-1}(p) - F^{-1}(p) \big)$ is asymptotically normal with mean zero and
variance $p/(1-p)/f^2\big(F^{-1}(p) \big)$. Thus, the standard deviation of
the empirical quantile of $p$, is approximately proportional to
$\sqrt{p(1-p)/I}$ which has its maximum value at $p=0.5$.
Thus, the critical values are computed with an approximate accuracy of
$0.5/\sqrt{I}$. With $I=10^8$, we have $0.5/\sqrt{I}=0.00005$, thus roughly indicating that the critical values in Tables~\ref{TBL:criticalvalues1} and \ref{TBL:criticalvalues2}
are accurate up to the fourth decimal point.

\section{Two illustrative examples} \label{section:03}
In this section, we consider two real-data applications 
to illustrate the proposed goodness-of-fit test for the Birnbaum-Saunders distribution based on the probability plot. 

\textbf{Example 1.} This example from \cite{Hsieh:1990} concerns the active repair times (in hours) for an airborne communications transceiver, and they are given by the following forty-six observations:
0.2, 0.3, 0.5, 0.5, 0.5, 0.5, 0.6, 0.6, 0.7, 0.7, 0.7, 0.8,
0.8, 1, 1, 1, 1, 1.1, 1.3, 1.5, 1.5, 1.5, 1.5, 2, 2, 2.2, 2.5,
2.7, 3, 3, 3.3, 3.3, 4, 4, 4.5, 4.7, 5, 5.4, 5.4, 7, 7.5, 8.8,
9, 10.3, 22, and 24.5.

\begin{figure}[t!]
\centering\includegraphics[width=6in]{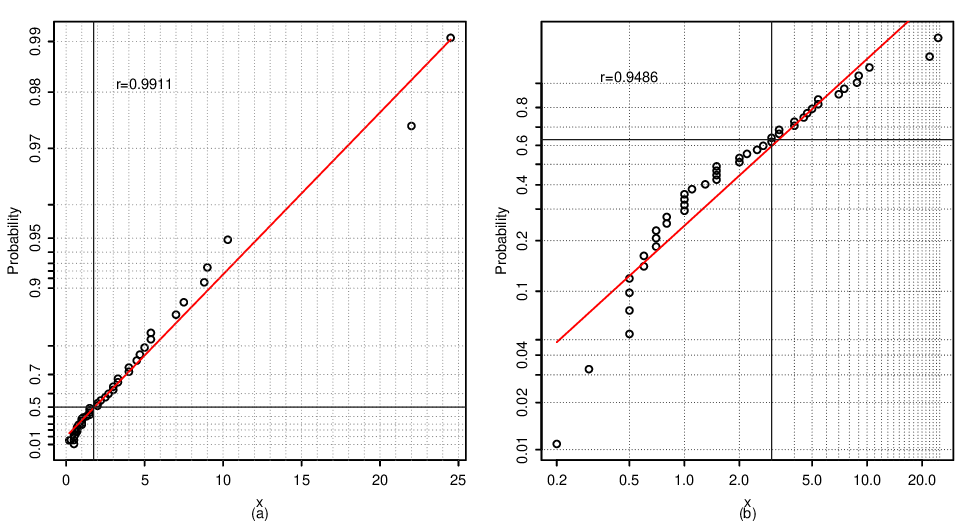}  
\caption{(a) Birnbaum-Saunders probability plot and (b) Weibull probability plot of the repair time data.\label{FIG:example}}
\end{figure}

The probability plots of the Birnbaum-Saunders and Weibull distributions for the repair time data are provided in Figure~\ref{FIG:example} (a) and (b), respectively. By utilizing the critical values for $n=46$ in Table~\ref{TBL:criticalvalues1}, we obtain $r_\gamma=0.9880$ at the significance level of $\gamma=25\%$ and $r_\gamma=0.9920$ at $\gamma=50\%$. Since the sample correlation coefficient from the Birnbaum-Saunders probability plot is $R=0.9911$, we conclude that the null hypothesis of the Birnbaum-Saunders distribution should be rejected at the significance level of 50\%, but not at
the level of 25\%.  In addition, to estimate the $p$-value using the table, we observe
that $R=0.9911$ falls between the critical values $r_\gamma=0.9880$ and $r_\gamma=0.9920$ at 
$\gamma=25\%$ and $\gamma=50\%$, respectively, indicating that that the $p$-value lies between 25\% and 50\%. By linearly interpolating between these values, we obtain the $p$-value of 44.2\%.

We also conducted the Weibull goodness-of-fit test for the same data through using the \texttt{weibullness} R package \citep{Park:2023b}. The sample correlation coefficient from the Weibull probability plot (Figure~\ref{FIG:example} (b)) is $R=0.9486$, and the corresponding $p$-value is remarkably low at only 1.69\%, thus resulting in the rejection of the null hypothesis of the Weibull distribution at the significance level of $\gamma=0.5\%$. 

\textbf{Example 2.} This experimental data set was originally from \cite{Smith/Naylor:1987} on the strength of glass fiber of length 15cm. The observations are given by the following forty-six observations: 0.37, 0.4, 0.7, 0.75, 0.8, 0.81, 0.83, 0.86, 0.92, 0.92, 0.94,
0.95, 0.98, 1.03, 1.06, 1.06, 1.08, 1.09, 1.1, 1.1, 1.13, 1.14, 
1.15, 1.17, 1.2, 1.2, 1.21, 1.22, 1.25, 1.28, 1.28, 1.29, 1.29, 
1.3, 1.35, 1.35, 1.37, 1.37, 1.38, 1.4, 1.4, 1.42, 1.43, 1.51,
1.53, and 1.61. 

\begin{figure}[t!]
\centering\includegraphics[width=6in]{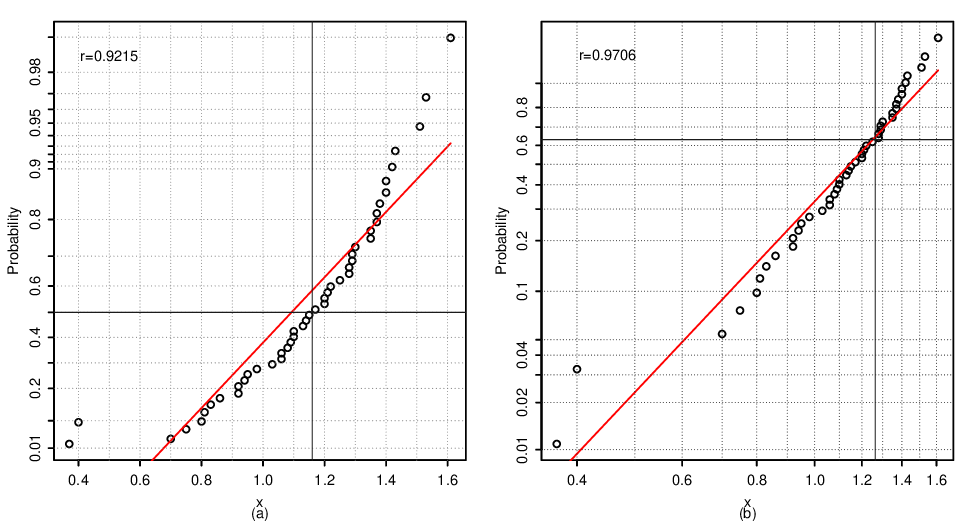}  
\caption{(a) Birnbaum-Saunders probability plot and (b) Weibull probability plot of the repair time data.
\label{FIG:example2}}
\end{figure}

The probability plots of the Birnbaum-Saunders and Weibull distributions for the repair time data are provided in Figure~\ref{FIG:example2} (a) and (b), respectively. 

The sample correlation coefficient from the Birnbaum-Saunders probability plot for this data set is
$R=0.9215$. By utilizing the critical value for $n=46$ from Table~\ref{TBL:criticalvalues1}, 
we have $r_\gamma=0.9639$ at the significance level of $\gamma=0.5\%$. 
As a result, the null hypothesis of the Birnbaum-Saunders distribution should be rejected at the significance level of 0.5\%, as the $p$-value of the goodness-of-fit test is smaller than 0.5\%. 

In addition, we conducted the Weibull goodness-of-fit test by employing the \texttt{weibullness} R package \citep{Park:2023b}. The sample correlation coefficient from the Weibull probability plot (Figure~\ref{FIG:example2} (b)) is $R=0.9706$, and the corresponding $p$-value is 9.326\%. Thus, we fail to reject the null hypothesis of the Weibull distribution for this data set.

\section{Concluding Remarks} \label{section:04}

In this paper, we have formulated a formal hypothesis test for Birnbaum-Saunders goodness-of-fit using the sample correlation coefficient as the test statistic. Given the challenge of deriving the null distribution of the sample correlation coefficient, we conducted extensive Monte Carlo simulations to acquire the empirical distribution from the Birnbaum-Saunders probability plot. Through these simulations, we derived critical values, corresponding significance levels, and p-values. We have provided critical values at specific significance levels, including 0.5\%, 1\%, 2\%, 2.5\%, 5\%, 7.5\%, 10\%, 12.5\%, 15\%, 20\%, 25\%, and 50\%. We believe that these values cater well to most practical applications.

\begin{figure}[t!]
\centering\includegraphics[width=6in]{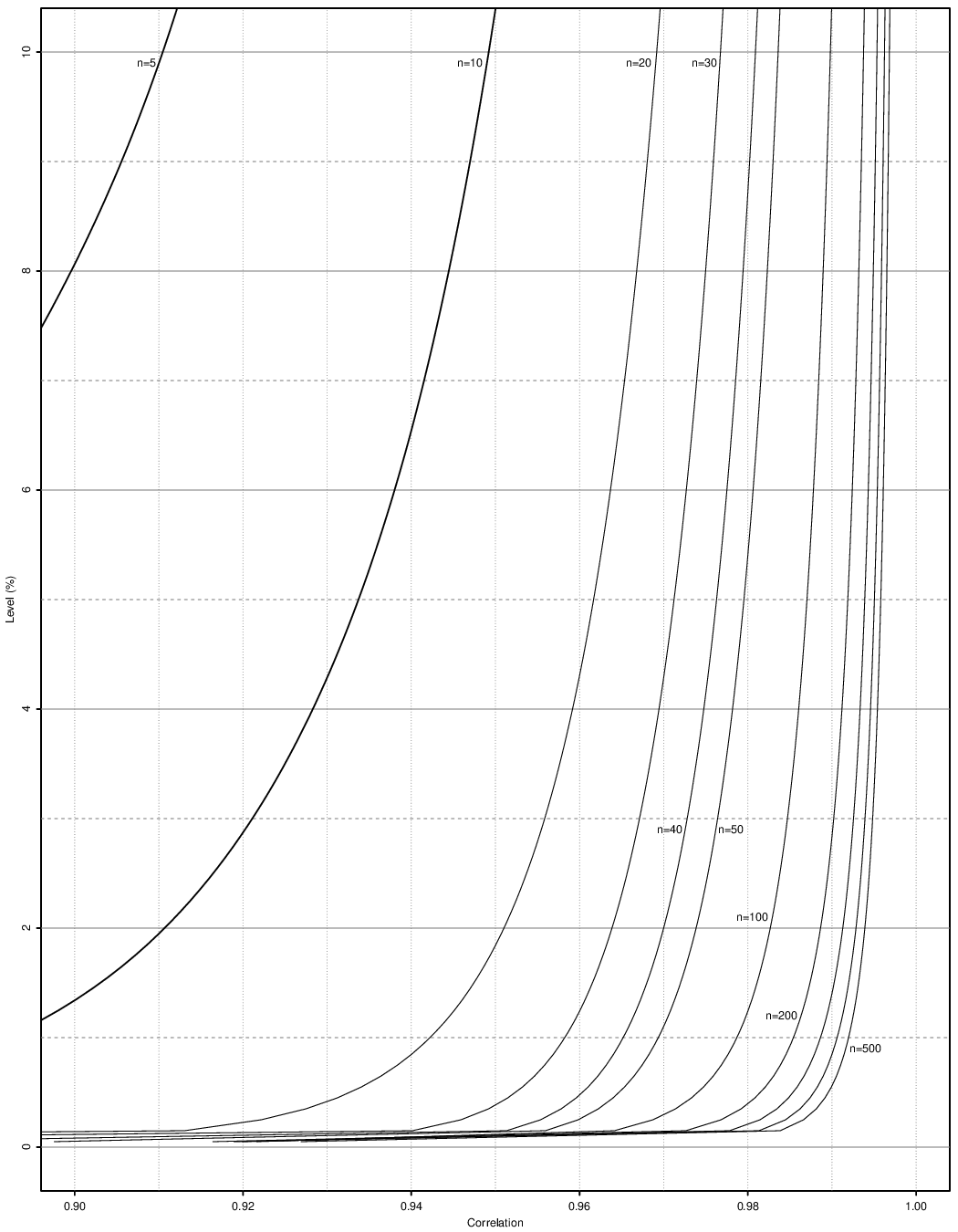}  
\caption{The zoom of the empirical CDF in Figure~1 with $\gamma$ up to 10\%\label{FIG:plot2}}
\end{figure}

It is worth noting that if critical values for different significance levels are needed, interpolation methods can be employed using the critical values in Tables \ref{TBL:criticalvalues1} and \ref{TBL:criticalvalues2}, or reference can be made to Figures~\ref{FIG:plot1} and \ref{FIG:plot2}. Alternatively, for more precise critical values, one could follow the approach outlined in the paper and perform additional Monte Carlo simulations following a similar pattern. Furthermore, we are actively working on developing a comprehensive R package library that will furnish all the critical values. This library will be made accessible on the author's personal webpage.



\section*{Acknowledgments}
This work was supported by the National Research Foundation of Korea (NRF) grants funded
by the Korea government (MSIT) (No.\ 2022R1A2C1091319 and RS-2023-00242528).

\bibliographystyle{apalike}
\bibliography{draftmw72.bib}

\clearpage
 \vspace*{2in}
\section*{\centering\Huge Appendix: Figures of the CDFs and Tables for the critical values}

\clearpage

\clearpage

\clearpage
\renewcommand{\baselinestretch}{1.000}
\begin{table}[t!]
\begin{center}
\caption{Critical values ($100 \times r_\gamma$) of the sample correlation coefficient
from the Birnbaum-Saunders probability plot.\label{TBL:criticalvalues1}}
\begin{scriptsize}
\begin{tabular}{clrrrrrrrrrrrr} 
\hline\hline \\[-2ex]
    && \multicolumn{12}{c}{Significance levels, $\gamma$} \\
$n$ && 0.5\%  & 1\%    & 2\%    & 2.5\%  & 5\%    & 7.5\%  & 10\%   & 12.5\% & 15\%   & 20\%   & 25\%   & 50\%  \\
\cline{1-1} \cline{3-14} \\[-2ex]
  3 && 72.08 & 74.45 & 77.27 & 78.31 & 82.03 & 84.67 & 86.89 & 88.84 & 90.44 & 92.73 & 94.25 & 97.87\\
  4 && 69.98 & 74.17 & 78.98 & 80.65 & 85.74 & 87.87 & 89.21 & 90.27 & 91.17 & 92.70 & 93.94 & 97.28\\
  5 && 75.08 & 78.89 & 82.50 & 83.67 & 87.42 & 89.65 & 91.06 & 91.97 & 92.65 & 93.71 & 94.58 & 97.25\\
  6 && 78.78 & 82.11 & 85.35 & 86.33 & 89.27 & 90.98 & 92.17 & 93.04 & 93.68 & 94.57 & 95.23 & 97.36\\
  7 && 81.88 & 84.63 & 87.36 & 88.23 & 90.74 & 92.13 & 93.10 & 93.82 & 94.38 & 95.19 & 95.75 & 97.51\\
  8 && 84.22 & 86.60 & 88.91 & 89.65 & 91.85 & 93.03 & 93.84 & 94.45 & 94.93 & 95.64 & 96.15 & 97.66\\
  9 && 86.02 & 88.12 & 90.15 & 90.79 & 92.71 & 93.75 & 94.45 & 94.97 & 95.38 & 96.00 & 96.45 & 97.79\\
 10 && 87.46 & 89.32 & 91.13 & 91.69 & 93.39 & 94.32 & 94.93 & 95.38 & 95.75 & 96.29 & 96.70 & 97.91\\
 11 && 88.36 & 90.02 & 91.65 & 92.16 & 93.70 & 94.56 & 95.13 & 95.56 & 95.91 & 96.43 & 96.82 & 97.99\\
 12 && 89.30 & 90.82 & 92.30 & 92.77 & 94.17 & 94.95 & 95.47 & 95.86 & 96.17 & 96.64 & 97.00 & 98.09\\
 13 && 90.11 & 91.50 & 92.85 & 93.28 & 94.56 & 95.28 & 95.75 & 96.11 & 96.39 & 96.83 & 97.16 & 98.18\\
 14 && 90.79 & 92.07 & 93.33 & 93.72 & 94.90 & 95.55 & 95.99 & 96.32 & 96.58 & 96.99 & 97.30 & 98.26\\
 15 && 91.37 & 92.57 & 93.73 & 94.10 & 95.19 & 95.79 & 96.20 & 96.51 & 96.75 & 97.13 & 97.42 & 98.33\\
 16 && 91.89 & 93.00 & 94.09 & 94.43 & 95.44 & 96.00 & 96.38 & 96.67 & 96.90 & 97.25 & 97.53 & 98.40\\
 17 && 92.34 & 93.38 & 94.40 & 94.72 & 95.66 & 96.18 & 96.54 & 96.81 & 97.03 & 97.37 & 97.63 & 98.46\\
 18 && 92.73 & 93.72 & 94.67 & 94.97 & 95.86 & 96.34 & 96.68 & 96.94 & 97.14 & 97.47 & 97.72 & 98.51\\
 19 && 93.09 & 94.02 & 94.92 & 95.19 & 96.03 & 96.49 & 96.81 & 97.05 & 97.25 & 97.56 & 97.80 & 98.56\\
 20 && 93.41 & 94.29 & 95.13 & 95.39 & 96.18 & 96.62 & 96.92 & 97.16 & 97.35 & 97.64 & 97.87 & 98.61\\
 21 && 93.70 & 94.53 & 95.32 & 95.57 & 96.31 & 96.73 & 97.03 & 97.25 & 97.43 & 97.72 & 97.94 & 98.65\\
 22 && 93.96 & 94.74 & 95.50 & 95.73 & 96.44 & 96.84 & 97.12 & 97.34 & 97.52 & 97.79 & 98.01 & 98.69\\
 23 && 94.19 & 94.94 & 95.65 & 95.87 & 96.55 & 96.94 & 97.21 & 97.42 & 97.59 & 97.86 & 98.07 & 98.73\\
 24 && 94.40 & 95.11 & 95.79 & 96.00 & 96.65 & 97.03 & 97.29 & 97.50 & 97.66 & 97.92 & 98.12 & 98.76\\
 25 && 94.59 & 95.27 & 95.91 & 96.12 & 96.75 & 97.11 & 97.37 & 97.56 & 97.73 & 97.98 & 98.17 & 98.80\\
 26 && 94.77 & 95.41 & 96.03 & 96.22 & 96.83 & 97.19 & 97.44 & 97.63 & 97.79 & 98.03 & 98.22 & 98.83\\
 27 && 94.93 & 95.54 & 96.13 & 96.32 & 96.92 & 97.26 & 97.50 & 97.69 & 97.84 & 98.08 & 98.27 & 98.86\\
 28 && 95.07 & 95.66 & 96.23 & 96.42 & 96.99 & 97.33 & 97.56 & 97.75 & 97.90 & 98.13 & 98.31 & 98.88\\
 29 && 95.20 & 95.76 & 96.32 & 96.50 & 97.06 & 97.39 & 97.62 & 97.80 & 97.95 & 98.18 & 98.35 & 98.91\\
 30 && 95.32 & 95.86 & 96.40 & 96.58 & 97.13 & 97.45 & 97.68 & 97.85 & 97.99 & 98.22 & 98.39 & 98.93\\
 31 && 95.43 & 95.95 & 96.48 & 96.65 & 97.19 & 97.51 & 97.73 & 97.90 & 98.04 & 98.26 & 98.42 & 98.96\\
 32 && 95.53 & 96.04 & 96.56 & 96.72 & 97.25 & 97.56 & 97.78 & 97.95 & 98.08 & 98.29 & 98.46 & 98.98\\
 33 && 95.62 & 96.12 & 96.62 & 96.79 & 97.31 & 97.61 & 97.82 & 97.99 & 98.12 & 98.33 & 98.49 & 99.00\\
 34 && 95.70 & 96.19 & 96.69 & 96.85 & 97.36 & 97.66 & 97.87 & 98.03 & 98.16 & 98.36 & 98.52 & 99.02\\
 35 && 95.78 & 96.26 & 96.75 & 96.91 & 97.41 & 97.71 & 97.91 & 98.07 & 98.20 & 98.40 & 98.55 & 99.04\\
 36 && 95.85 & 96.32 & 96.81 & 96.97 & 97.46 & 97.75 & 97.95 & 98.11 & 98.23 & 98.43 & 98.58 & 99.06\\
 37 && 95.92 & 96.38 & 96.87 & 97.02 & 97.51 & 97.79 & 97.99 & 98.14 & 98.26 & 98.46 & 98.60 & 99.07\\
 38 && 95.99 & 96.44 & 96.92 & 97.07 & 97.55 & 97.83 & 98.02 & 98.17 & 98.30 & 98.48 & 98.63 & 99.09\\
 39 && 96.05 & 96.50 & 96.97 & 97.12 & 97.59 & 97.87 & 98.06 & 98.21 & 98.33 & 98.51 & 98.65 & 99.11\\
 40 && 96.10 & 96.55 & 97.02 & 97.17 & 97.63 & 97.90 & 98.09 & 98.24 & 98.35 & 98.54 & 98.68 & 99.12\\
 41 && 96.16 & 96.60 & 97.06 & 97.21 & 97.67 & 97.94 & 98.12 & 98.27 & 98.38 & 98.56 & 98.70 & 99.14\\
 42 && 96.21 & 96.65 & 97.11 & 97.25 & 97.71 & 97.97 & 98.16 & 98.30 & 98.41 & 98.59 & 98.72 & 99.15\\
 43 && 96.25 & 96.70 & 97.15 & 97.29 & 97.75 & 98.00 & 98.18 & 98.32 & 98.43 & 98.61 & 98.74 & 99.16\\
 44 && 96.30 & 96.74 & 97.19 & 97.33 & 97.78 & 98.04 & 98.21 & 98.35 & 98.46 & 98.63 & 98.76 & 99.18\\
 45 && 96.34 & 96.78 & 97.23 & 97.37 & 97.81 & 98.06 & 98.24 & 98.37 & 98.48 & 98.65 & 98.78 & 99.19\\
 46 && 96.39 & 96.82 & 97.26 & 97.41 & 97.84 & 98.09 & 98.27 & 98.40 & 98.51 & 98.67 & 98.80 & 99.20\\
 47 && 96.42 & 96.86 & 97.30 & 97.44 & 97.87 & 98.12 & 98.29 & 98.42 & 98.53 & 98.69 & 98.82 & 99.21\\
 48 && 96.46 & 96.90 & 97.33 & 97.47 & 97.90 & 98.15 & 98.32 & 98.44 & 98.55 & 98.71 & 98.83 & 99.22\\
 49 && 96.50 & 96.93 & 97.37 & 97.51 & 97.93 & 98.17 & 98.34 & 98.47 & 98.57 & 98.73 & 98.85 & 99.23\\
 50 && 96.53 & 96.97 & 97.40 & 97.54 & 97.96 & 98.20 & 98.36 & 98.49 & 98.59 & 98.75 & 98.87 & 99.24\\
 55 && 96.69 & 97.12 & 97.55 & 97.68 & 98.08 & 98.31 & 98.46 & 98.58 & 98.68 & 98.83 & 98.94 & 99.29\\
 60 && 96.83 & 97.25 & 97.67 & 97.80 & 98.19 & 98.41 & 98.55 & 98.67 & 98.76 & 98.90 & 99.00 & 99.33\\
 65 && 96.94 & 97.37 & 97.78 & 97.91 & 98.28 & 98.49 & 98.63 & 98.74 & 98.82 & 98.96 & 99.06 & 99.37\\
 70 && 97.04 & 97.47 & 97.87 & 98.00 & 98.36 & 98.56 & 98.70 & 98.80 & 98.88 & 99.01 & 99.10 & 99.40\\
 75 && 97.13 & 97.56 & 97.96 & 98.08 & 98.44 & 98.63 & 98.76 & 98.86 & 98.94 & 99.06 & 99.15 & 99.43\\
 80 && 97.21 & 97.64 & 98.04 & 98.15 & 98.50 & 98.69 & 98.81 & 98.91 & 98.98 & 99.10 & 99.19 & 99.46\\
 85 && 97.28 & 97.71 & 98.10 & 98.22 & 98.56 & 98.74 & 98.86 & 98.95 & 99.02 & 99.14 & 99.22 & 99.48\\
 90 && 97.35 & 97.78 & 98.17 & 98.28 & 98.61 & 98.79 & 98.91 & 98.99 & 99.06 & 99.17 & 99.25 & 99.50\\
 95 && 97.42 & 97.84 & 98.23 & 98.34 & 98.66 & 98.83 & 98.95 & 99.03 & 99.10 & 99.20 & 99.28 & 99.52\\
100 && 97.47 & 97.90 & 98.28 & 98.39 & 98.71 & 98.87 & 98.98 & 99.06 & 99.13 & 99.23 & 99.30 & 99.54\\
\hline\hline  \\
\end{tabular}
\end{scriptsize}
\end{center}
\end{table}

\clearpage
\renewcommand{\baselinestretch}{1.000}
\begin{table}[t!]
\begin{center}
\caption{Critical values ($100 \times r_\gamma$) of the sample correlation coefficient
from the Birnbaum-Saunders probability plot.\label{TBL:criticalvalues2}}
\begin{scriptsize}
\begin{tabular}{clrrrrrrrrrrrr} 
\hline\hline \\[-2ex]
    && \multicolumn{12}{c}{Significance levels, $\gamma$} \\
$n$ && 0.5\%  & 1\%    & 2\%    & 2.5\%  & 5\%    & 7.5\%  & 10\%   & 12.5\% & 15\%   & 20\%   & 25\%   & 50\%  \\
\cline{1-1} \cline{3-14} \\[-2ex]
110 && 97.58 & 98.00 & 98.37 & 98.48 & 98.79 & 98.94 & 99.05 & 99.13 & 99.19 & 99.28 & 99.35 & 99.57\\
120 && 97.68 & 98.09 & 98.46 & 98.56 & 98.85 & 99.00 & 99.10 & 99.18 & 99.24 & 99.32 & 99.39 & 99.59\\
130 && 97.76 & 98.17 & 98.53 & 98.63 & 98.91 & 99.06 & 99.15 & 99.22 & 99.28 & 99.36 & 99.42 & 99.61\\
140 && 97.84 & 98.24 & 98.59 & 98.69 & 98.97 & 99.11 & 99.20 & 99.26 & 99.31 & 99.39 & 99.45 & 99.64\\
150 && 97.91 & 98.31 & 98.65 & 98.75 & 99.01 & 99.15 & 99.23 & 99.30 & 99.35 & 99.42 & 99.48 & 99.65\\
160 && 97.98 & 98.37 & 98.70 & 98.80 & 99.06 & 99.18 & 99.27 & 99.33 & 99.38 & 99.45 & 99.50 & 99.67\\
170 && 98.04 & 98.42 & 98.75 & 98.84 & 99.09 & 99.22 & 99.30 & 99.36 & 99.40 & 99.47 & 99.53 & 99.68\\
180 && 98.10 & 98.48 & 98.80 & 98.89 & 99.13 & 99.25 & 99.33 & 99.38 & 99.43 & 99.50 & 99.55 & 99.70\\
190 && 98.16 & 98.52 & 98.84 & 98.92 & 99.16 & 99.28 & 99.35 & 99.41 & 99.45 & 99.51 & 99.56 & 99.71\\
200 && 98.21 & 98.57 & 98.87 & 98.96 & 99.19 & 99.30 & 99.38 & 99.43 & 99.47 & 99.53 & 99.58 & 99.72\\
210 && 98.26 & 98.61 & 98.91 & 98.99 & 99.22 & 99.33 & 99.40 & 99.45 & 99.49 & 99.55 & 99.59 & 99.73\\
220 && 98.30 & 98.65 & 98.94 & 99.02 & 99.24 & 99.35 & 99.42 & 99.47 & 99.51 & 99.56 & 99.61 & 99.74\\
230 && 98.34 & 98.68 & 98.97 & 99.05 & 99.27 & 99.37 & 99.44 & 99.48 & 99.52 & 99.58 & 99.62 & 99.75\\
240 && 98.38 & 98.71 & 99.00 & 99.08 & 99.29 & 99.39 & 99.45 & 99.50 & 99.54 & 99.59 & 99.63 & 99.75\\
250 && 98.42 & 98.75 & 99.02 & 99.10 & 99.31 & 99.41 & 99.47 & 99.51 & 99.55 & 99.60 & 99.64 & 99.76\\
260 && 98.46 & 98.78 & 99.05 & 99.13 & 99.33 & 99.42 & 99.48 & 99.53 & 99.56 & 99.61 & 99.65 & 99.77\\
270 && 98.49 & 98.81 & 99.07 & 99.15 & 99.34 & 99.44 & 99.50 & 99.54 & 99.58 & 99.63 & 99.66 & 99.77\\
280 && 98.52 & 98.83 & 99.10 & 99.17 & 99.36 & 99.45 & 99.51 & 99.55 & 99.59 & 99.64 & 99.67 & 99.78\\
290 && 98.56 & 98.86 & 99.12 & 99.19 & 99.38 & 99.47 & 99.52 & 99.57 & 99.60 & 99.64 & 99.68 & 99.79\\
300 && 98.59 & 98.88 & 99.14 & 99.21 & 99.39 & 99.48 & 99.54 & 99.58 & 99.61 & 99.65 & 99.69 & 99.79\\
310 && 98.62 & 98.91 & 99.16 & 99.22 & 99.41 & 99.49 & 99.55 & 99.59 & 99.62 & 99.66 & 99.70 & 99.80\\
320 && 98.64 & 98.93 & 99.17 & 99.24 & 99.42 & 99.50 & 99.56 & 99.60 & 99.63 & 99.67 & 99.70 & 99.80\\
330 && 98.67 & 98.95 & 99.19 & 99.26 & 99.43 & 99.51 & 99.57 & 99.60 & 99.63 & 99.68 & 99.71 & 99.81\\
340 && 98.70 & 98.97 & 99.21 & 99.27 & 99.44 & 99.53 & 99.58 & 99.61 & 99.64 & 99.68 & 99.72 & 99.81\\
350 && 98.72 & 98.99 & 99.22 & 99.29 & 99.46 & 99.54 & 99.59 & 99.62 & 99.65 & 99.69 & 99.72 & 99.81\\
360 && 98.74 & 99.01 & 99.24 & 99.30 & 99.47 & 99.54 & 99.59 & 99.63 & 99.66 & 99.70 & 99.73 & 99.82\\
370 && 98.77 & 99.03 & 99.25 & 99.31 & 99.48 & 99.55 & 99.60 & 99.64 & 99.66 & 99.70 & 99.73 & 99.82\\
380 && 98.79 & 99.05 & 99.27 & 99.33 & 99.49 & 99.56 & 99.61 & 99.64 & 99.67 & 99.71 & 99.74 & 99.83\\
390 && 98.81 & 99.06 & 99.28 & 99.34 & 99.50 & 99.57 & 99.62 & 99.65 & 99.68 & 99.72 & 99.74 & 99.83\\
400 && 98.83 & 99.08 & 99.29 & 99.35 & 99.51 & 99.58 & 99.62 & 99.66 & 99.68 & 99.72 & 99.75 & 99.83\\
410 && 98.85 & 99.09 & 99.30 & 99.36 & 99.51 & 99.59 & 99.63 & 99.66 & 99.69 & 99.73 & 99.75 & 99.84\\
420 && 98.86 & 99.11 & 99.32 & 99.37 & 99.52 & 99.59 & 99.64 & 99.67 & 99.69 & 99.73 & 99.76 & 99.84\\
430 && 98.88 & 99.12 & 99.33 & 99.38 & 99.53 & 99.60 & 99.64 & 99.68 & 99.70 & 99.74 & 99.76 & 99.84\\
440 && 98.90 & 99.14 & 99.34 & 99.39 & 99.54 & 99.61 & 99.65 & 99.68 & 99.71 & 99.74 & 99.77 & 99.84\\
450 && 98.92 & 99.15 & 99.35 & 99.40 & 99.55 & 99.61 & 99.66 & 99.69 & 99.71 & 99.75 & 99.77 & 99.85\\
460 && 98.93 & 99.16 & 99.36 & 99.41 & 99.55 & 99.62 & 99.66 & 99.69 & 99.72 & 99.75 & 99.77 & 99.85\\
470 && 98.95 & 99.18 & 99.37 & 99.42 & 99.56 & 99.63 & 99.67 & 99.70 & 99.72 & 99.75 & 99.78 & 99.85\\
480 && 98.96 & 99.19 & 99.38 & 99.43 & 99.57 & 99.63 & 99.67 & 99.70 & 99.72 & 99.76 & 99.78 & 99.85\\
490 && 98.98 & 99.20 & 99.39 & 99.44 & 99.57 & 99.64 & 99.68 & 99.71 & 99.73 & 99.76 & 99.79 & 99.86\\
500 && 98.99 & 99.21 & 99.40 & 99.45 & 99.58 & 99.64 & 99.68 & 99.71 & 99.73 & 99.76 & 99.79 & 99.86\\
550 && 99.06 & 99.26 & 99.44 & 99.48 & 99.61 & 99.67 & 99.71 & 99.73 & 99.75 & 99.78 & 99.80 & 99.87 \\
600 && 99.12 & 99.31 & 99.47 & 99.52 & 99.63 & 99.69 & 99.72 & 99.75 & 99.77 & 99.80 & 99.82 & 99.88 \\
650 && 99.17 & 99.35 & 99.50 & 99.55 & 99.66 & 99.71 & 99.74 & 99.76 & 99.78 & 99.81 & 99.83 & 99.88 \\
700 && 99.21 & 99.38 & 99.53 & 99.57 & 99.68 & 99.72 & 99.76 & 99.78 & 99.79 & 99.82 & 99.84 & 99.89 \\
750 && 99.25 & 99.42 & 99.55 & 99.59 & 99.69 & 99.74 & 99.77 & 99.79 & 99.80 & 99.83 & 99.85 & 99.90 \\
800 && 99.29 & 99.44 & 99.58 & 99.61 & 99.71 & 99.75 & 99.78 & 99.80 & 99.81 & 99.84 & 99.85 & 99.90 \\
850 && 99.32 & 99.47 & 99.59 & 99.63 & 99.72 & 99.76 & 99.79 & 99.81 & 99.82 & 99.84 & 99.86 & 99.91 \\
900 && 99.35 & 99.49 & 99.61 & 99.65 & 99.73 & 99.77 & 99.80 & 99.82 & 99.83 & 99.85 & 99.87 & 99.91 \\
950 && 99.38 & 99.51 & 99.63 & 99.66 & 99.74 & 99.78 & 99.81 & 99.82 & 99.84 & 99.86 & 99.87 & 99.91 \\
1000&& 99.40 & 99.53 & 99.64 & 99.67 & 99.75 & 99.79 & 99.81 & 99.83 & 99.84 & 99.86 & 99.88 & 99.92 \\
\hline\hline  \\
\end{tabular}
\end{scriptsize}
\end{center}
\end{table}

\end{document}